\documentclass[fleqn,usenatbib]{mnras}

\usepackage{newtxtext,newtxmath}

\usepackage[T1]{fontenc}
\usepackage{ae,aecompl}

\usepackage{graphicx}	
\usepackage{amsmath}	
\usepackage{amssymb}	

\newcommand{\GB}{{\sc GREENBURST}}	
\newcommand{\rateunits}{$\text{day}^{-1} \text{sky}^{-1}$}
\newcommand{\dmunits}{$\text{pc}\; \text{cm}^{-3}$}

\title[Initial results from GREENBURST]{Initial results from a realtime FRB search with the GBT}

\author[Agarwal et al.] {Devansh Agarwal$^{1,2}$\thanks{da0017@mix.wvu.edu}, D.~R. Lorimer$^{1,2}$, M.P.~Surnis$^{1,2}$,
X.~Pei$^{3,4}$, A.~Karastergiou$^{5,6,7}$,
\newauthor
G.~Golpayegani$^{1,2}$, 
D.~Werthimer$^{10}$,
J.~Cobb$^{10}$,
M.A.~McLaughlin$^{1,2}$, 
S.~White$^{8}$,  
\newauthor
W.~Armour$^{9}$, 
D.H.E.~MacMahon$^{10}$,
A.P.V.~Siemion$^{10,11,12}$ and
G.~Foster$^{5}$.\\
$^{1}$West Virginia University, Department of Physics and Astronomy, P. O. Box 6315, Morgantown, WV, USA\\
$^{2}$Center for Gravitational Waves and Cosmology, West Virginia University, Chestnut Ridge Research Building, Morgantown, WV, USA\\
$^{3}$Xinjiang Astronomical Observatory, Chinese Academy of Sciences, Urumqi, Xinjiang 830011, China\\
$^{4}$University of Chinese Academy of Sciences, Beijing 100049, China\\
$^{5}$University of Oxford, Sub-Department of Astrophysics, Denys Wilkinson Building, Keble Road, Oxford, OX1 3RH, United Kingdom\\
$^{6}$Physics Department, University of the Western Cape, Cape Town 7535, South Africa\\
$^{7}$Department of Physics and Electronics, Rhodes University, PO Box 94, Grahamstown 6140, South Africa\\
$^{8}$Green Bank Observatory, P.O. Box 2, Green Bank, WV 24944, USA\\
$^{9}$OeRC, Department of Engineering Science, University of Oxford, Keble Road, Oxford OX1 3QG, UK\\
$^{10}$Department of Astronomy, University of California, Berkeley, 501 Campbell Hall \#3411, Berkeley, CA 94720, USA\\
$^{11}$Radboud University, Nijmegen, 6525 HP, the Netherlands\\
$^{12}$SETI Institute, Mountain View, CA 94043, USA\\
}

\date{Accepted XXX. Received YYY; in original form ZZZ}

\pubyear{2020}

\begin{document}
\label{firstpage}
\pagerange{\pageref{firstpage}--\pageref{lastpage}}
\maketitle

\begin{abstract}
We present the data analysis pipeline, commissioning observations and initial results from the GREENBURST fast radio burst (FRB) detection system on the Robert C. Byrd Green Bank Telescope (GBT)
previously described by Surnis et al. which uses the 21~cm receiver observing  commensally with other projects. The pipeline makes use of a state-of-the-art deep learning classifier
to winnow down the very large number of false positive single-pulse candidates that mostly result from
radio frequency interference. In our observations totalling 156.5 days so far, we have detected individual pulses from 20 known radio pulsars which provide an excellent verification of the system performance. We also demonstrate, through blind injection analyses, that our pipeline is complete down to a signal-to-noise threshold of 12. Depending on the observing mode, this translates to peak flux sensitivities in the range 0.14--0.89~Jy. Although no FRBs have been detected to date, we have used our results to update the analysis of Lawrence et al. to constrain the FRB all-sky rate to be $1140^{+200}_{-180}$ per day above a peak flux density of 1~Jy. We also constrain the source count index $\alpha=0.83\pm0.06$ which indicates that the source count distribution is substantially flatter
than expected from a Euclidean distribution of standard candles (where $\alpha=1.5$). We discuss this result in the context of the FRB redshift and luminosity distributions. Finally, we make predictions for  detection rates with GREENBURST, as well as other ongoing and planned FRB experiments.
\end{abstract}




\section{Introduction}
Fast Radio bursts (FRBs) are enigmatic astrophysical objects that burst for millisecond durations with flux densities of the order of a few Jansky that were first discovered by \citet{Lorimer2007}. FRBs show the characteristic inverse frequency squared sweep in observing frequency, which is quantified by the dispersion measure (DM). Their DMs are substantially larger than those expected from the Milky Way in the direction of detection, indicating their extragalactic nature. To date, $\sim$110 FRBs have been reported\footnote{\url{http://frbcat.org}} \citep[see][and references therein]{frbcatalog}, and their origins are still unclear. Out of all the FRBs discovered so far, of special interest are those which show repeat bursts. The first repeater, FRB~121102, \citep{Spitler2014} was localised to a host galaxy by \citet{Chatterjee2017}. FRB~171019 was found to have repeat bursts with $\sim350$ times smaller fluence as compared to the first detection \citep{Shannon2018, Kumar19} and 18 repeaters were also recently reported \citep{abb+19b, Andersen2019, Fonseca2020}. Very recently, a 16.34-day periodicity from the repeating FRB~180916.J0158+65 \citep{Dongzi2020} and a possible 159-day periodicity for FRB~121102 \citep{2020arXiv200303596R} were announced.

Detection of FRBs requires data at radio frequency to be de-dispersed at many trial DM values. For each DM, all the frequencies are added to form a time series which is then searched using matched filters to find bursts above a certain threshold. With the help of Graphics Processing Units (GPUs), it is now possible to perform such searches in real time \citep{Magro2011, Barsdell2012, artemis2015, Adamek2019}. Inspired by the capabilities of real-time processing which has been successfully implemented at Parkes \citep[see][for recent commensal discoveries]{Osowski2019}, many radio telescopes around the globe are deploying commensal search backends to enable serendipitous discoveries of FRBs. A few examples include: \textsc{realfast} \citep{Law2018} at the Very Large Array, the \textsc{craft} survey with the Australian Square Kilometre Array Pathfinder (ASKAP) telescope \citep{Macquart2010}, \textsc{alfaburst} at the Arecibo observatory \citep{Chennamangalam2015,Foster2018} and GBTrans using the 20~m telescope at Green Bank \citep{Golpayegani2019}. With such backends, a copy of the data from the receiver is de-dispersed and searched for FRBs. Real-time detection of FRBs is required for prompt follow-up at other wavelengths that might provide valuable insights towards understanding the underlying emission mechanisms and possible progenitors.

In this paper we present the results from  3756~hours on sky from the commensal backend at the 110~m Robert C. Byrd Green Bank Telescope (GBT). We henceforth refer to this system as  \GB. This paper is organised as follows. We first describe and summarise the system description and detail the FRB search pipeline in \S\ref{sec:search_pipeline} followed by benchmarks of our pipeline in \S\ref{sec:benchmark}. In \S\ref{sec:results} we present results from our commensal observations and constraints on FRB rates. In  \S\ref{sec:discussion}, we discuss the consequences of our results in terms of FRB source counts and predictions for ongoing future experiments. Finally, in \S\ref{sec:conclusion}, we present our conclusions.

\section{Search Pipeline}

\label{sec:search_pipeline}
\begin{figure*}
    \centering
    \includegraphics[width=\textwidth]{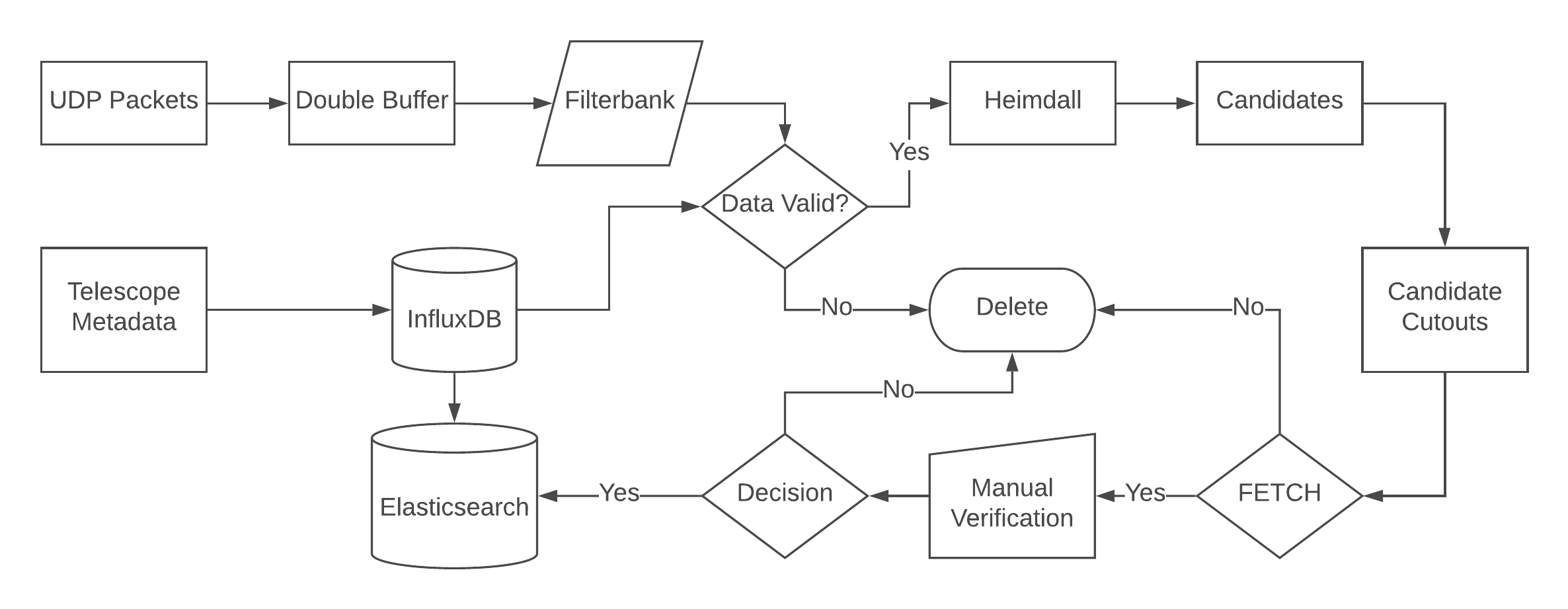}
    \caption{Schematic depiction of the detection pipeline. Data through ethernet arrives as user datagram protocol (UDP) packets. Using a double buffer system, data from the UDP packets are rearranged and written in filterbank format. In parallel, all the telescope metadata are saved in the influx database at 1~s intervals. Once a filterbank file is written, data validity is checked (see text for details). Valid data are searched with \textsc{heimdall}. Candidates are then parsed through \textsc{FETCH}, and positively labelled candidates are sent for visual inspection. A condensed version of telescope metadata and the candidates is saved in elasticsearch for future reference.}
    \label{fig:01_pipeline}
\end{figure*}

The system description is detailed in \citet{Surnis2019} and is summarised here. Using a dedicated directional coupler designed and built at the observatory, we obtain a copy of the signal from the L-band (21~cm) receiver. This signal is then digitised using a field programmable gate array on board the {\sc setiburst} backend \citep{Chennamangalam2017} and sampled every 256~$\mu$s with 8-bit precision. The resulting data stream consists of 4096 channels spanning a 960~MHz bandwidth at a central frequency of 1440~MHz. A unique property of this system is that even when the L-band receiver is  not in the primary focus, it still is illuminated by a large part of the dish. As a result, it can be used commensally with observations at other frequencies. 

Fig.~\ref{fig:01_pipeline} details our search and verification pipeline. The digitised data are transported over an ethernet connection to a dedicated computer which processes and stores the data as binary files in {\tt filterbank} format \citep{Lorimer2000}.  The filterbank files contain 16 chunks of $2^{17}$ samples corresponding to 10,000~\dmunits~DM delay along with an overlapping chunk from the last file. The overlap ensures that no transient events are missed due to data being split between two files. In parallel, the telescope metadata, which includes the receiver turret angle, telescope pointing altitude and azimuth, and observing project IDs are recorded at a cadence of one second in influxDB\footnote{\url{https://www.influxdata.com}}. This serves as a high-resolution short term storage database, where the metadata are saved for seven days. 

\begin{figure*}
    \centering
    \includegraphics[width=\textwidth]{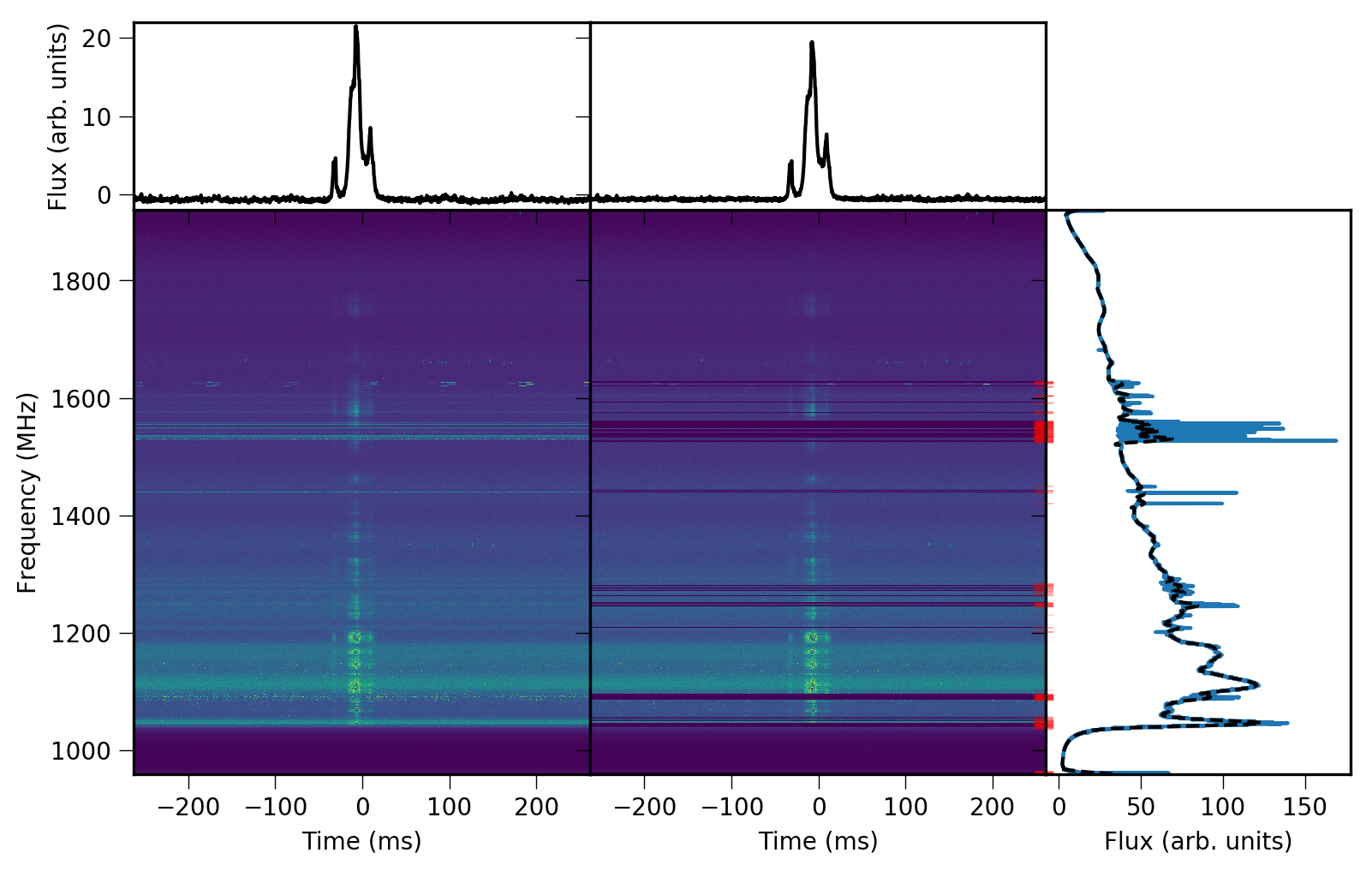}
    \caption{Radio frequency interference clipping using the Savitzky--Golay filter. The bottom left, and middle plots show the raw and cleaned de-dispersed spectrum of a single from PSR~B0329+54. In the bottom right panel, the raw bandpass is shown in blue, while the smoothed bandpass from the filter is shown by the black dashed lines. The red lines mark the flagged channels. The top left and middle plots show the frequency integrated profile of the single pulse.}
    \label{fig:savgol}
\end{figure*}

Once a filterbank file is written, data validity is checked using metadata from influxDB. The data are considered invalid if any of the following conditions are met.
\begin{itemize}
    \item The receiver turret is unlocked. This typically happens when the observer changes the receiver in focus.
    \item The turret angle is between 160$^\circ$ and  220$^\circ$. At these angles the GBT primary focus feed structure blocks the receiver's field of view.
    \item The primary focus receiver is extended due to the same reason as above.
\end{itemize}
If the data are valid, we first excise radio frequency interference (RFI) from affected channels using the following method. All the time samples are added to form a bandpass of the data. The bandpass is smoothed using a Savitzky--Golay filter. Here we use a running window of 61 data samples and fit a second-order polynomial to obtain a smooth bandpass. The measured and smooth bandpass are subtracted from one another. Through empirical investigations with preliminary data, we found that a good RFI excision procedure is to use this subtraction result and flag any channels which differ from the smooth bandpass by more than five counts\footnote{Here we use the term ``count'' to refer to an intensity value quantized in the range 0--255. Five counts corresponds to $\sim6$ times the root mean square value of the data.}.  Both the window size and the difference threshold were determined empirically. Fig.~\ref{fig:savgol} shows the profile of PSR~B0329+54, before and after masking bad channels. In parallel, a coarse version of the telescope metadata is computed by binning by time spent by the telescope in each 1$^\circ \times$1$^\circ$ patch in the Galactic latitude-longitude grid. The metadata are subsequently used to generate sky coverage maps and rate calculations of FRBs described below.

We use {\sc heimdall}\footnote{\url{https://sourceforge.net/projects/heimdall-astro}} along with the bad channel flags to search for pulses in the range $10 \leq$ DM $\leq 10,000$~\dmunits, and smoothing over $[2^0,2^1,...,2^7]$ adjacent samples spanning widths in the range 256~$\mu$s---32.768~ms above a signal to noise ratio (S/N) of 8. The candidates above the S/N threshold are then classified as either RFI or an astronomical transient using model \texttt{a} of the artificial neural network {\sc FETCH} \citep{2019arXiv190206343A}. Candidates labelled as positives are then sorted into two categories: Galactic and extragalactic. We do this by computing the expected DM contribution in the direction of observation by integrating the electron density by both NE2001 and YMW16 models out to 25~kpc. The smaller of the two DM estimates is chosen as the Galactic DM in that direction. In case the candidate DM is Galactic, the position and DM are matched with the ATNF pulsar catalogue \citep{Manchester2005} to verify if the candidate is a known source. If the source is unknown or the DM is larger than the Galactic DM, the candidates are marked for manual verification. Positively marked candidates are stored in the elasticsearch\footnote{\url{https://www.elastic.co/elasticsearch}} database.

\section{Pipeline Benchmarks}
\label{sec:benchmark}

To assess the completeness of our pipeline, we injected fake FRBs with various observational parameters and run the complete pipeline as detailed in \S\ref{sec:search_pipeline}. Based on the results from our pipeline we compute several metrics to quantify the pipeline's ability to detect FRBs.

\subsection{Blind FRB injections}
\begin{table}
 \caption{Distributions of FRBs injected for benchmarking the pipeline}
 \label{tab:frb_injection}
 \centering
 \begin{tabular}{ll}
  \hline
  Parameter name & Distribution \\
  \hline
  Signal-to-noise ratio & Uniform(6,100) \\
  Pulse width & Uniform(0.5, 26)~ms\\
  Spectral index & Uniform(--3,3)\\
  Scattering time & Uniform(0.256,6.5)~ms\\
  Number of scintillation patches & Log-Uniform(--3, 2)\\
  \hline
 \end{tabular}
\end{table}

To inject FRBs, we first randomly select filterbank files from the observations on a single day (MJD 58728). On this date, all the data were acquired using the L-band receiver. The parameters of the injected FRB distribution are summarised in Table~\ref{tab:frb_injection}. For each injection, first a random start time in the file is chosen such that there is enough data to fully inject the dispersion delay. Then, Gaussian-shaped profiles are created for each channel with standard deviation
\begin{equation}
    w = \sqrt{t_\text{samp}^2 + t_\text{DM}^2 + w_\text{int}^2}.
\end{equation}
Here $t_\text{samp}=256~\mu$s is the sampling interval, $t_\text{DM}$ is the dispersion smearing (the delay due to dispersion across a  channel bandwidth) and $w_\text{int}$ is the intrinsic pulse width drawn from a uniform distribution between 0.5 and 26~ms. This profile is then convolved with an exponential function of the form $e^{-t/\tau}/\tau$, where $\tau$ is randomly drawn from a uniform distribution of 0.256 and 6.5~ms, to add scattering the the profile. The lower limit is 0.256~ms and not zero because of the $1/\tau$ normalisation factor and the upper limit of 6.5~ms such that the resultant widths of the scattered FRBs are similar to the observed population. These profiles are then scaled with the spectral index by multiplying with $(F/F_\text{ref})^\gamma$. Here $F$ is the channel frequency, $F_\text{ref}$ is the reference frequency of 1400~MHz and $\gamma$ is the spectral index.
Scintillation is added to the data by modulating the spectra using the positive half of a cosine function. The number of such patches are drawn from a log-normal distribution of mean --3 and standard deviation of 2. The above parameters lead to $\sim10$~\% of FRBs with a patchy spectral structure. To add scintillation we create an envelope where $N_s$ is the number of bright patches which is multiplied with the pulse. The envelope, 
\begin{equation}
    E = \cos \bigg[ 2 \pi N_s \Big(\frac{F}{F_\text{ref}}\Big)^2 + \phi \bigg],
\end{equation}
is generated with $\phi$ being a random phase in the range 0 to $2\pi$ drawn from a uniform distribution. $E>0$ values are then multiplied with $S$ to simulate scintillation. The parameters from the above-described distributions are drawn and injected using the publicly available code \texttt{injectfrb}\footnote{\url{https://github.com/liamconnor/injectfrb}}.
To create realistic bright FRBs, as our the data are 8-bit unsigned integers, for cases where the profile intensity exceeds the dynamic range the values are wrapped around the maximum value of 255. This is done because the FPGA wraps the numbers  exceeding the dynamic range instead of clipping them at the maximum value. An instance of this can be viewed in Fig.~\ref{fig:savgol} where dark blue patches can be seen within the dynamic spectrum of the pulsar.

\subsection{Evaluation Metrics}

\begin{figure}
    \centering
    \includegraphics[width=\columnwidth]{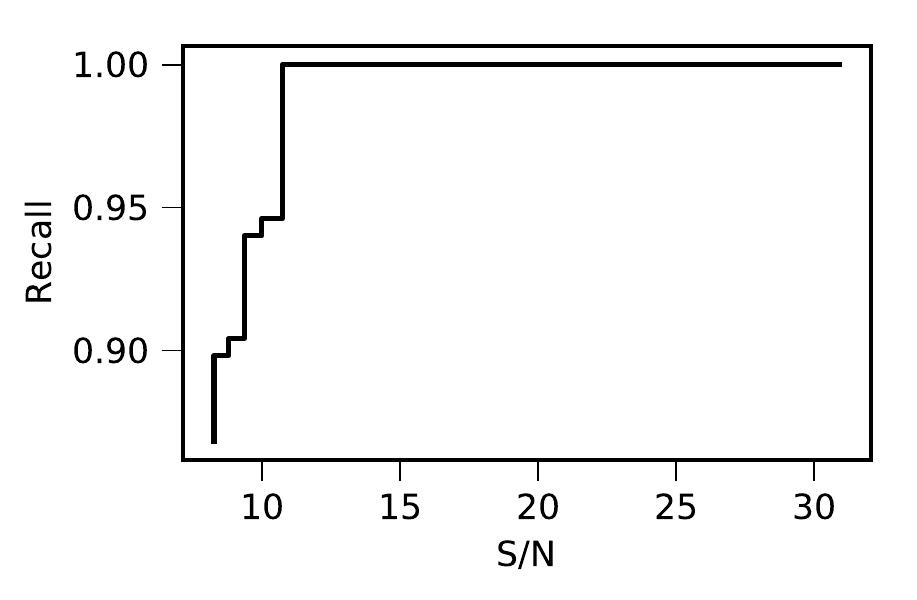}
    \caption{The parameteric recall curve. The ordinate and abscissa correspond to the injected S/N and the recall respectively. The curve indicates that our pipeline is able to recover all the injected FRBs with S/N $\gtrsim$12.}
    \label{fig:pr}
\end{figure}

To quantify the performance of our pipeline, we calculate what is known as ``recall'' \citep{encyclopediaofmachinelearninganddatamining2017} which is simply the ratio of the number of recovered FRBs to the total number of injected FRBs. While there are other metrics like accuracy and precision, their calculation involves the number of false positives which themselves depend on the RFI environment at the time of the observations and the performance of the RFI mitigation algorithms. We restrict our evaluation to the recovery of injected FRBs, and hence, we chose to evaluate using recall. To extract deeper insights than traditional recall, we here define a parameter weighted recall which we call parametric recall (PR). For this analysis we inject $\sim$1200 FRBs and we chose S/N as the  parameter. Then, injected data are binned with respect to the parameter such that each bin has an equal number of points and the recall is calculated for each bin. PR can also be understood as the first moment of a distribution of recall over the given parameter. In this framework, we have
\begin{equation}
    \text{PR} =  \frac{\sum_{i=0}^{N_{\text{bins}}} \text{Recall}_i \, \mathcal{P}_i}{\sum_{i=0}^{N_{\text{bins}}} \mathcal{P}_i}. 
\end{equation}
Here, $\mathcal{P}_i$ and $\text{Recall}_i$ is the mean $\mathcal{P}$ and the recall of the $i^{\rm th}$ bin. The maximum value for the PR is unity, i.e.~the pipeline found all FRBs at all injected S/N values. In case where the pipeline misses FRBs at high S/N the PR would be penalised more resulting in lower overall score. Hence PR is a better measure of performance as compared to traditional recall.

Fig.~\ref{fig:pr} shows the PR for the injected S/N as parameter ($\mathcal{P}$). As can be seen from the plot, the pipeline is able to recover all  events above a S/N $\sim$12. The PR  from the above stated curve is 0.95. We inspected the candidates missed by the pipeline injected between a S/N of 8 to 12. All the candidates missed are due to the presence of strong RFI near the signal. In future, we plan to implement more sophisticated RFI mitigation algorithms to prevent achieve a lower S/N threshold with 100\% reliability.

\section{Results}
\label{sec:results}
\GB~started commensal observations on MJD 58587 (2019-03-14) and, as of MJD 58917 (2020-03-09), has observed for 156.5 days. While the backend has been operational for 330 days, only $\sim$50\% of the available time has been spent on sky. This because of several factors that govern the validity of the data such as the telescope down time for maintenance, availability and observer's choice of the receiver (see \S\ref{sec:search_pipeline} for details).

\begin{figure*}
   \centering
    \includegraphics[width=\textwidth]{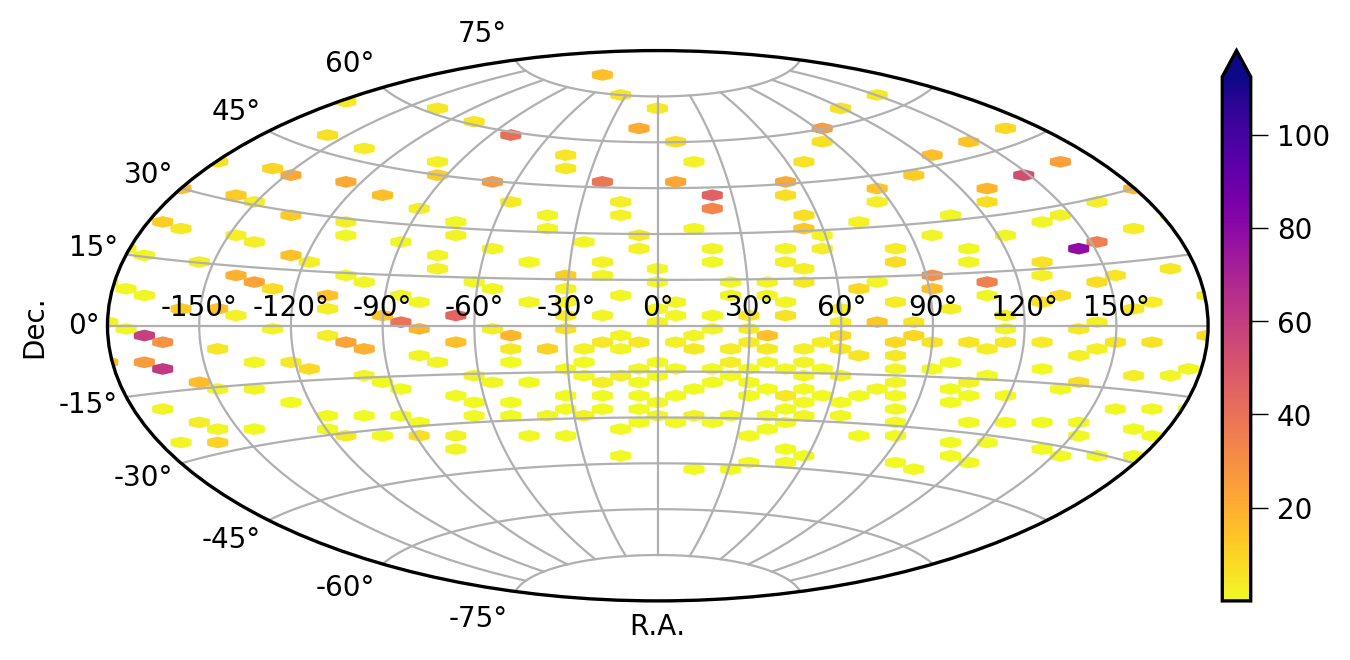}
    \caption{Sky coverage during commensal observations. The figure shows the converge as 36~deg$^2$ hexagonal bins in the sky as an equatorial projection. as the respective axes. The color bar denotes the total hours spent in each bin by all turret positions.}
    \label{fig:hours_spent}
\end{figure*}

Fig.~\ref{fig:hours_spent} shows the sky coverage during this time in equatorial coordinates. The hexagons show $6^{\circ} \times 6^{\circ}$ area with colour bar representing the hours spent in the region.

\begin{table}
 \caption{GREENBURST observational summary to date. From left to right, we list the receiver in prime focus, the turret angle relative to the L-band receiver, the time spent on sky with that receiver, the instantaneous solid angle covered 
 $(\Omega)$ and the sensitivity as evaluated from the blind injection analysis (see \S 3.1).}
 \label{tab:100hours}
 \begin{tabular}{lrrrr }
 \hline
Receiver & Turret & Observation  & $\Omega$  & Sensitivity\\
         & Angle & Time & $\times 10^{-2}$& \\
         & ($^\circ$) & (hr) &  (sr) & (Jy) \\
\hline
L-band   & 0            & 2194      & 3.12 & 0.14       \\
X-band   & 260          & 615       & 3.33 & 0.89      \\
C-band   & 60           & 556       & 3.19 & 0.25      \\
Ku-band  & 100          & 210       & 3.40 & 0.80      \\
MUSTANG  & 300          & 181       & 3.26 & 0.26     \\
\hline
\end{tabular}
\end{table}

Table~\ref{tab:100hours} shows the time spent, solid angle and sensitivity at each turret position for ${\rm S/N}=12$. The sensitivity shown here is slightly different when compared to the numbers we reported earlier \citep{Surnis2019} where we assumed a bandwidth of 960~MHz for the calculation and a S/N threshold of 12. Soon after the backend became functional, due to the presence of RFI, it was decided to always have the notch filter  which blocks frequencies in the range 1.25--1.35~GHz in place. This filter is only taken out by the observer (primarily for pulsar/FRB observations). Along with the notch filter, we routinely flag $\sim$10\% of the total band band reducing our bandwidth to 760~MHz. The beam solid angle, $\Omega \approx 1.33\, \text{FWHM}^2$, where FWHM is the full width at half maximum and is taken from \citet{Surnis2019}. 

\begin{table}
 \caption{Known pulsars detected by GREENBURST during commensal observations. $N_{\text{pulses}}$ is the number of single pulses detected, S/N$_{\text{max}}$ is the max S/N detected for the corresponding pulsar. DM and the $S_{1400}$ is the dispersion measure and the mean flux density at 1400 MHz respectively from the ATNF pulsar catalogue.}
 \label{tab:pulsar_sp}
 \begin{tabular}{lrrrr}
 \hline
Pulsar   & \multicolumn{1}{c}{DM}   & \multicolumn{1}{c}{$N_{\text{pulses}}$}& \multicolumn{1}{c}{S/N$_{\text{max}}$} & \multicolumn{1}{c}{$S_{1400}$}\\
         & (\dmunits)      & & & (mJy)  \\
\hline
B0329+54    &  26.76   &  113  &  195  &  203  \\
J0426+4933  &  85.00   &  1    &  17   &  0.19    \\
B0450--18    &  39.90   &  423  &  77   &  16.8   \\
B0818--13    &  40.94   &  258  &  115  &  6    \\
B0919+06    &   27.29  &  2     &  14 & 10      \\
B1508+55    &  19.62   &  49   &  29   &  8    \\
B1702--19    &  22.91   &  316  &  80   &  9.3    \\
B1718--35    &  496.00  &  4    &  11   &  16.8   \\
B1745--20A   &  219.40  &  21   &  13   &  0.37    \\
B1804--08    &  112.38  &  102  &  30   &  18.2   \\
B1822--09    &  19.38   &  71   &  137  &  10.2   \\
B1933+16    &  158.52  &  408  &  262  &  57.8   \\
B1937+21    &  71.02   &  14   &  17   &  15.2   \\
B1946+35    &  129.37  &  125  &  133  &  8.3    \\
B2021+51    &  22.55   &  26   &  51   &  27   \\
B2035+36    &  93.56   &  2    &  42   &  0.8    \\
B2111+46    &  141.26  &  28   &  99   &  19   \\
B2154+40    &  71.12   &  58   &  75   &  17   \\
B2217+47    &  43.50   &  90   &  73   &  3    \\
B2310+42    &  17.28   &  43   &  43   &  15   \\
\hline
\end{tabular}
\end{table}

\begin{figure*}
    \centering
    \includegraphics[width=0.9\textwidth]{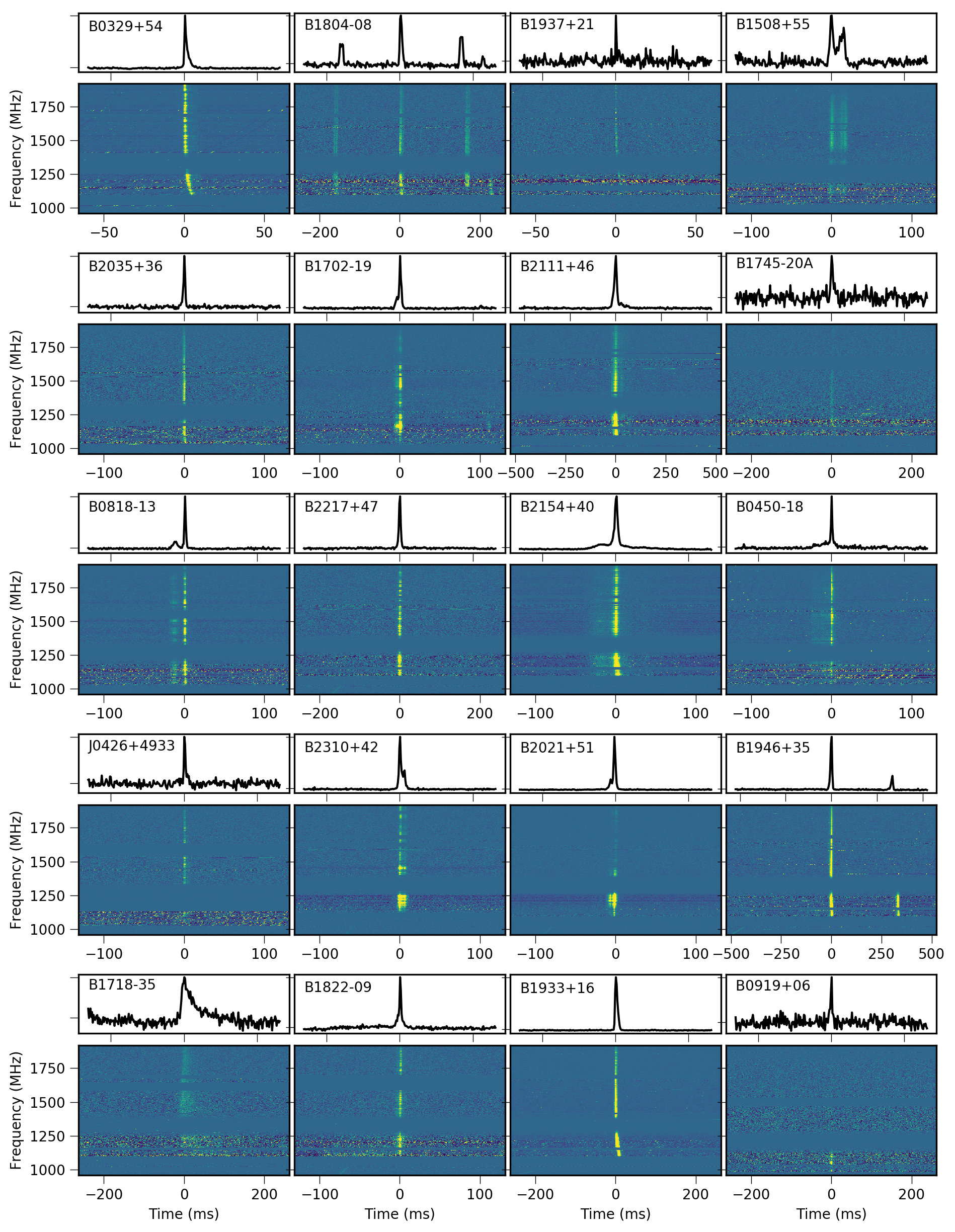}
    \caption{Brightest single pulses from various pulsars dedispersed at their detection DM. The figure shows the waterfall plot and frequency integrated time profile of the brightest pulses from 20 pulsars listed in Table~\ref{tab:pulsar_sp}. Pulsars are marked in the top left corner in each plot respectively.}
    \label{fig:pulsars_sp}
\end{figure*}

During observations so far, we detected 2153 single pulses from 20 pulsars. Table \ref{tab:pulsar_sp} shows the number of single pulses observed from each pulsar. Fig.~\ref{fig:pulsars_sp} shows the waterfall plot and frequency integrated time profile of the brightest single pulse from each pulsar. The pulsars in the figure are de-dispersed at the detection DM and at the DM of the pulsar. The two DMs are often different because the detection DM is a sample from the coarser grid of trial DMs used for the search. The presence of RFI and zero DM subtraction also contributes towards the difference between the DMs. As a result in some cases the effects of residual dispersion can be seen. In case of PSR~B1804-08 we can see three single pulses from the pulsar (the fourth pulse is narrow band RFI). For PSR~B1946+35 the burst near $\sim 300$~ms is also RFI. 

\section{Discussion}
\label{sec:discussion}

\subsection{Time to first GREENBURST detection}

So far, we have observed for 156.5 days and detected no FRBs. To check whether our non-detection is anticipated, we first use previous estimates of the all-sky rate of FRBs
\begin{equation}
    \mathcal{R}(\mathcal{S}) = \mathcal{R}_0 \, \left(\frac{\mathcal{S}}{\rm Jy}\right)^{-\alpha},
\end{equation}
where $\mathcal{R}_0$ is the reference rate and $\alpha$ is the source count index from the log~$N$--log~$S$ relation. In their analysis, \citet{Lawrence_2017} found ${\cal R}=587_{-305}^{+337}$ events per day
per sky and $\alpha=0.91 \pm 0.34$ where the uncertainties indicate the 95\% confidence interval. Using these parameters,
we estimate the waiting time to discover an FRB, $\mathcal{W} = 1/\mathcal{R}\Omega$ where $\Omega$ is the beam solid angle. Using the rates from \citet{Lawrence_2017}, we find $\mathcal{W}=532^{+1042}_{-184}$~days for the first detection. This is significantly larger than our present observing time. 

\subsection{The all-sky FRB rate}

We now use our null result to update the non-homogeneous Poisson process framework developed by \citet{Lawrence_2017} to find revised estimates
$\mathcal{R}$ as well as the source count index $\alpha$ of FRBs by taking into account both the detections and non-detections. 
We implemented the analysis described by 
\citet{Lawrence_2017} using the information from 12 surveys which included 15 detections. We extend this analysis by adding 13 surveys (including this work) with 32 FRBs. We extend the datasets of \citet{Lorimer2007} by including FRB~010312 \citep{Zhang2019} which is the second FRB in the original data set, and \citet{thornton2013} by including FRB~110214 \citep{Petroff2018} which was found by processing the remaining  0.5\% of the HTRU survey. We include 23 FRBs from ASKAP  \citep{Shannon2018, Qiu2019, Bhandari2019, Agarwal2019}. We also include 8 FRBs from the Parkes telesope  \citep{Bhandari2017,Osowski2019}. We also incorporate various surveys reporting non-detections \citep{Men2019,Golpayegani2019,Madison2019}. In this work, since each turret position has a different sensitivity and observing time, observations at position has been added as a different survey (see Table~\ref{tab:100hours} for details).

For this analysis, we exclude the FRBs from CHIME and UTMOST as they were carried out at different observing frequencies and have non-Gaussian beamshapes which are currently not incoporated into the framework. We also exclude several other surveys which have reported non-detections but were carried out in different frequency bands.

\begin{figure}
    \centering
    \includegraphics[width=\columnwidth]{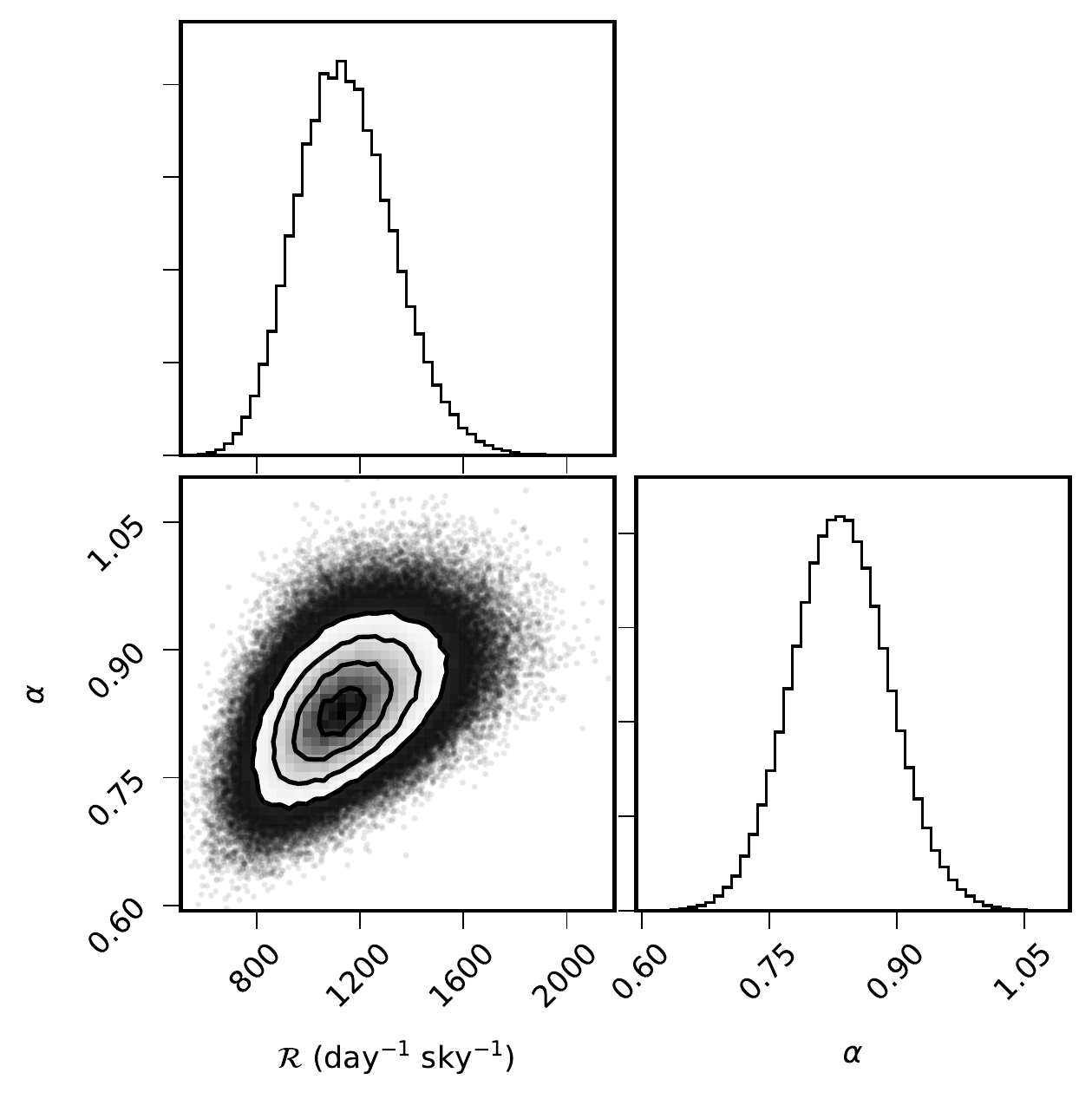}
    \caption{Joint and marginalized probability density functions for the FRB rate, ${\mathcal R}$, and source count index, $\alpha$, that were returned by our implementation of the Bayesian framework developed by Lawrence et al.~(2017).}
    \label{fig:mcmc_rate}
\end{figure}

We implement the likelihood formalism of \citet{Lawrence_2017} and use Markov Chain Monte Carlo (MCMC) simulation to obtain distributions of $\mathcal{R}_0$ and $\alpha$. We implement the MCMC using the {\sc EMCEE}\footnote{\url{https://github.com/dfm/emcee}} framework \citep{emcee} with a uniform prior of $\alpha$ and a log-uniform prior on $\mathcal{R}_0$. The resultant posterior distributions for $\log(\mathcal{R}_0)$ and $\alpha$ are shown in Fig.~\ref{fig:mcmc_rate}. From this analysis, we infer the FRB rate
\begin{equation}
\mathcal{R} = 1140^{+200}_{-180}\; \left(\frac{S}{\rm Jy}\right)^{-0.83\pm0.06}\, \text{day}^{-1} \text{sky}^{-1}.
\label{eq:rate}
\end{equation}
Here the quoted uncertainties corresponding to 95\% confidence intervals. We find a higher rate for the FRBs above 1~Jy, as compared to the \citeauthor{Lawrence_2017} which was $587^{+337}_{-315}$~\rateunits, however, the error regions with both the estimates overlap. Our source count index distribution is shallower than the \citeauthor{Lawrence_2017} value of $0.91\pm0.34$ but lies within their predicted ranges.

Based on this revised event rate, we predict that (for observations exclusively at L-band), \GB~will require a further $270^{+65}_{-89}$~days to make its first detection. As can been seen from Table \ref{tab:100hours}, L-Band is in focus for only $\sim$65\% of the total on sky time. Hence a more realistic estimate for the time to first detection is $365^{+88}_{-120}$~days.

\subsection{Detection rate forecasts for other surveys}

\begin{table}
 \caption{FRB detection rate predictions for various telescopes. From left to right, for each experiment, we list the telescope's field of view (FOV), the observing bandwidth ($\Delta \nu$), the centre frequency ($F_\text{centre}$) and the system equivalent flux density (SEFD) as well as the predicted rate ($R$).}
 \label{tab:forecast}
 \begin{tabular}{lrrrrr}
 \hline
Telescope   & \multicolumn{1}{c}{FOV}   & \multicolumn{1}{c}{$\Delta \nu$} & \multicolumn{1}{c}{$F_\text{centre}$} &  \multicolumn{1}{c}{SEFD} &  \multicolumn{1}{c}{$R$} \\
              & (deg$^2$) & (MHz) & (MHz)   & (Jy) & (day$^{-1}$) \\
\hline
CHIME         & 200       & 400   & 600     & 45   & 9 $\pm$ 2    \\
HIRAX         & 56        & 400   & 600     & 6    & 10 $\pm$ 3  \\
CHORD         & 130       & 1200  & 900     & 9    & 4 $\pm$ 1    \\
Northern Cross & 350       & 16    & 408     & 95   & 2 $\pm$ 1    \\
\hline
\end{tabular}
\end{table}
Using our estimates from Eq.~\ref{eq:rate} we compute expected FRB rates for experiments planned with four telescopes: CHIME \citep{Kaspi2018}, CHORD \citep{Vanderlinde_2019}, Northern Cross \citep{Locatelli_2020} and HIRAX \citep{Newburgh2016}. To estimate the rate for each survey, we compute the minimum flux density using the radiometer equation assuming a S/N threshold of 10. For experiments at frequencies outside of L-band, we assume a flat spectral index (i.e.~no scaling of ${\cal R}$ with frequency). For CHIME and CHORD, the system equivalent flux density
\begin{equation}
{\rm SEFD} = \frac{T_\text{rec} + T_\text{sky}}{G},
\end{equation}
where $T_\text{rec}$ and $T_\text{sky}$ are the receiver and the sky temperatures, respectively, and $G$ is the antenna gain. $T_\text{sky}$ is estimated using an average sky temperature of 34~K and a spectral index of --2.6 at a reference frequency of 408~MHz \citep{Haslam_1982}. The results from these calculations are shown in Table~\ref{tab:forecast}. We also cross-check our results against published detections from the UTMOST telescope, where \citet{Caleb2017} report  three FRBs from a 180-day survey. Our prediction for UTMOST over that time period is slightly higher (5 $\pm$ 1 detections) but does not account for the fact that a fraction of the UTMOST survey was conducted at reduced sensitivity \citep{Caleb2017}. Our forecasted rates for the other surveys are very promising and highlight the impact that these surveys will have on future constraints of the all-sky FRB rate.

\subsection{Source count index}

\begin{figure}
   \centering
    \includegraphics[width=\columnwidth]{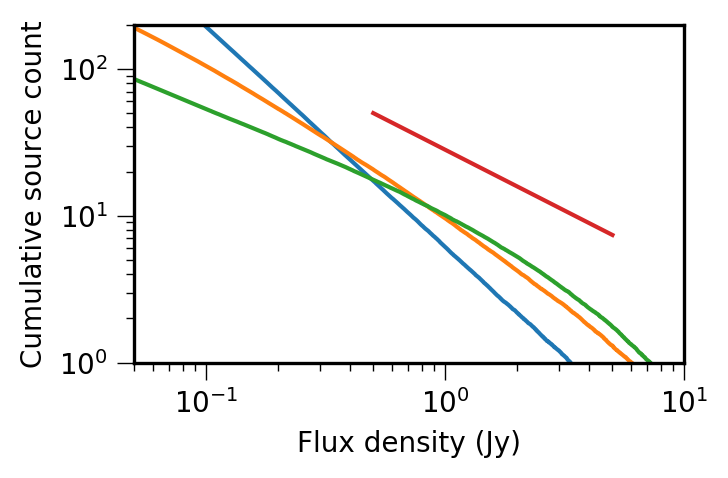}
    \caption{Model FRB source counts under the assumptions of: uniform distribution of standard candles in Euclidean space (blue line); a log-normal luminosity distribution uniformly distributed in comoving volume (green line); a log-normal luminosity function with redshift distribution following the cosmic star formation rate \citep[Eq.~15 of][orange line]{Madau2014}. The isolated red line shows the slope obtained from our analysis ($\alpha=0.83$) for comparison.}
    \label{fig:logNlogS}
\end{figure}

Our update of the FRB event rate favors a shallower slope $\alpha=0.83$ compared to the expectation from a population of standard candles uniformly distributed in Euclidean space for which $\alpha=1.5$. These lines are shown in the log~$N$--log~$S$ plane in Fig.~\ref{fig:logNlogS} and are clearly inconsistent with one another. Although detailed analyses of FRB source counts can be found elsewhere \citep[see, e.g.,][]{Macquart2017,2018MNRAS.480.4211M,2019MNRAS.483.1342J}, to show what can be learned from future discoveries, it is instructive to place our result in context
of two different cosmological models. These are also shown in Fig.~\ref{fig:logNlogS} and were computed using a simple Monte Carlo simulation in which FRBs were drawn from a population uniformly distributed in comoving volume (green line in the figure) and from a redshift distribution that follows the cosmic star formation history
\citep[see, Eq.~15 of][]{Madau2014}. From the corresponding redshift distributions, luminosity distances were computed for each Monte Carlo sample. In both these cases, the luminosities were assumed to be log--normal in form with a standard deviation (in log space) that is 2\% of the mean. The mean luminosity was set somewhat arbitrarily for the purpose of these simulations to be $10^{26}$~W. Flux densities were then computed which resulted in the corresponding cumulative curves. 

These models were chosen merely to demonstrate that the impact of these assumptions is to naturally flatten the slope of the source count function from the Euclidean value to something that more closely resembles what is observed. Also shown in these simulations is a steepening of the slopes at higher flux density values. Our analysis in section 5.2 does not account for a possible change in $\alpha$ across the log~$N$--log~$S$ plane. In their analysis of Parkes and ASKAP detections, where they considered fluence rather than flux density, \citet{2019MNRAS.483.1342J} also found a steepening of the slope at higher fluence values which they suggested could be due to a change in the redshift distribution of the sources. Further analyses of the source count function are definitely required and likely to result in significant insights, particularly from CHIME where a sample of $\sim 700$ FRBs are eagerly anticipated \citep{2020ApJ...891L...6F}.

\section{Conclusions}
\label{sec:conclusion}
In this paper, we present results from the first 156.5~days of commensal FRB searches at the GBT. We use a GPU accelerated single-pulse search pipeline and classify candidates using a deep learning-based algorithm. Our pipeline searches and classifies candidates in real time and logs the relevant telescope metadata using several databases. 
During our observations, we detected over 2000 single pulses from 20 pulsars which helped to validate our pipeline. We also carried out blind injection analysis of the data and find that we could categorically detect all FRBs with S/N greater than 12. 

Our null result is in line with the FRB rate estimates by \citet{2017ApJ...841L..12B}. We update the analysis and report a rate of $1140^{+200}_{-180}$~\rateunits~and a shallow source count index of $0.83\pm 0.03$ above a peak flux of 1~Jy. We estimate a further year of observations is required to result in \GB~FRB detections. With a stable observing system now in place, we anticipate being able to go well beyond this expectation through continued commensal observing. Our revised FRB rate shows that emerging and ongoing experiments have excellent prospects to discover a very large sample of FRBs in the coming years. Through a  Monte Carlo simulation, we show that studies of the source counts of FRBs using this sample will provide significant insights into the luminosity and redshift distributions of FRBs.

\section*{Acknowledgements}

The Green Bank Observatory is a facility of the National Science Foundation (NSF) operated under cooperative agreement by Associated Universities, Inc. We acknowledge support from the NSF awards AAG-1616042, OIA-1458952 and PHY-1430284. 

\bibliographystyle{mnras}
\bibliography{greenburst} 
\bsp	
\label{lastpage}
\end{document}